## Journal of Geophysical Research: Space Physics

# What controls the local time extent of flux transfer events?

S. E. Milan[1,2], S. M. Imber[1], J. A. Carter[1], M.-T. Walach[1], and B. Hubert[3]

[1]Department of Physics and Astronomy, University of Leicester, Leicester, UK, [2]Birkeland Centre for Space Science, University of Bergen, Bergen, Norway, [3]Laboratory of Planetary and Atmospheric Physics, University of Liège, Liège, Belgium

**Abstract** Flux transfer events (FTEs) are the manifestation of bursty and/or patchy magnetic reconnection at the magnetopause. We compare two sequences of the ionospheric signatures of flux transfer events observed in global auroral imagery and coherent ionospheric radar measurements. Both sequences were observed during very similar seasonal and interplanetary magnetic field (IMF) conditions, though with differing solar wind speed. A key observation is that the signatures differed considerably in their local time extent. The two periods are 26 August 1998, when the IMF had components $B_Z \approx -10$ nT and $B_Y \approx 9$ nT and the solar wind speed was $V_X \approx 650$ km s$^{-1}$, and 31 August 2005, IMF $B_Z \approx -7$ nT, $B_Y \approx 17$ nT, and $V_X \approx 380$ km s$^{-1}$. In the first case, the reconnection rate was estimated to be near 160 kV, and the FTE signatures extended across at least 7 h of magnetic local time (MLT) of the dayside polar cap boundary. In the second, a reconnection rate close to 80 kV was estimated, and the FTEs had a MLT extent of roughly 2 h. We discuss the ramifications of these differences for solar wind-magnetosphere coupling.

## 1. Introduction

Magnetic reconnection occurring at the low-latitude magnetopause when the interplanetary magnetic field (IMF) has a southward component is the primary mechanism by which dynamics are driven within the magnetosphere-ionosphere system [e.g., *Dungey*, 1961; *Cowley and Lockwood*, 1992; *Milan et al.*, 2007]. However, the factors that control the reconnection rate are poorly understood. Numerous observational studies have concluded that the reconnection rate is proportional to the motional electric field of the solar wind, with an IMF clock angle factor that maximizes when the IMF is directed southward [e.g., *Milan et al.*, 2012, and references therein], though it is not clear why this should be [e.g., *Borovsky and Birn*, 2014]. Numerical simulations have investigated the dependence of the reconnection rate on upstream conditions, including solar wind Mach number and IMF clock angle [e.g., *Borovsky et al.*, 2008; *Ouellette et al.*, 2010]. The clock angle of the IMF modifies the magnetic shear angle between the magnetosheath and magnetopause, leading to predictions regarding the reconnection geometries available near the subsolar point and at higher latitudes [e.g., *Trattner et al.*, 2012, and references therein]. However, the solar wind velocity control of the location of reconnection has received less attention. In this paper we investigate the width of the ionospheric convection throat in the ionosphere, which is related to the extent of the reconnection X line on the magnetopause, and the factors which may control this.

*Baker et al.* [1997], *Pinnock and Rodger* [2001], *Milan et al.* [2003], *Hubert et al.* [2006], and *Chisham et al.* [2008] have outlined methods for determining the extent and rate of dayside reconnection by measuring the location and speed of ionospheric convective flows across the dayside polar cap boundary or open/closed field line boundary. However, no systematic study has been conducted. In this paper we use signatures of transient reconnection to estimate and compare the width of the convection throat during two intervals with similar IMF strength and orientation, but differing solar wind speed.

Considerable evidence has accumulated that even when the IMF is steadily southward magnetic reconnection can occur in an episodic and/or patchy manner. For instance, bipolar magnetic signatures in the magnetopause-normal direction have been interpreted as bundles of newly reconnected flux tubes [e.g., *Russell and Elphic*, 1978, 1979; *Haerendel et al.*, 1978; *Dunlop et al.*, 2005; *Fear et al.*, 2008, 2009], commonly known as flux transfer events (FTEs). Quasiperiodic features in the dayside auroral zone, including poleward moving auroral forms (PMAFs) and ionospheric convection flow bursts or poleward moving radar auroral forms (PMRAFs), have been interpreted as the ionospheric counterpart of FTEs [e.g., *Sandholt et al.*, 1986, 1992;

*Lockwood et al.*, 1989, 2001; *Pinnock et al.*, 1993, 1995; *Fasel*, 1995; *Moen et al.*, 1995; *Provan et al.*, 1998; *Neudegg et al.*, 1999; *Milan et al.*, 1999a, 1999b, 2000; *Wild et al.*, 2001; *Marchaudon et al.*, 2004].

Several studies have investigated the position and local time extent of transient reconnection at the magnetopause and in the ionosphere. FTE signatures can be seen across the dayside magnetopause [e.g., *Rijnbeek et al.*, 1984; *Elphic and Southwood*, 1987; *Kawano and Russell*, 1996, 1997], and *Lockwood et al.* [1995] used the local time distribution observed by *Rijnbeek et al.* [1984] to argue that FTEs could contribute the whole dayside reconnection rate. However, magnetopause FTEs are reconnected flux tubes which can have been transported far from their origin by the time they are observed, making it difficult to localize the reconnection X line. Having said this, observations of reconnection jets at the flanks of the magnetosphere indicate that the X line can indeed extend far from noon [e.g., *Phan et al.*, 2006].

Similarly, studies of the distribution of FTE signatures in the ionosphere [e.g., *Lockwood and Davis*, 1996; *Provan et al.*, 1999; *Milan et al.*, 2000; *Wild et al.*, 2005; *Fear et al.*, 2008, 2009, 2012; *Carter et al.*, 2015] have demonstrated that reconnection signatures can be observed over a wide range of magnetic local times (MLT) and that there appears to be an IMF $B_Y$ control of the location. Such studies have also shown that the FTE signatures can differ considerably in local time extent, from small scales of order a few 100 km [e.g., *Oksavik et al.*, 2004, 2005] to larger scales encompassing 4–5 h of MLT [e.g., *Lockwood et al.*, 1990]. These studies have not been able, however, to determine what controls the instantaneous width of the reconnection X line.

*Milan et al.* [2000] reported the first PMAF signatures in near-global images of the auroras taken from space. Combined observations from the Ultraviolet Instrument (UVI) on board the Polar spacecraft [*Torr et al.*, 1995] and measurements of the associated ionospheric flow by the Super Dual Auroral Radar Network (SuperDARN) [*Chisham et al.*, 2007] on 26 August 1998 demonstrated that FTEs can extend across at least 7 h of MLT. This is even larger than suggested by previous studies, which employed ground-based optical imagery with limited fields of view.

The current paper presents only the second example of the auroral signature of FTEs in global auroral imagery. These auroral signatures were observed on 31 August 2005 in the Southern Hemisphere by the Imager for Magnetopause-to-Aurora Global Exploration (IMAGE) Wideband Imaging Camera of the far ultraviolet instrument suite (FUV/WIC) [*Mende et al.*, 2000a, 2000b], while the associated ionospheric flows were measured in the Northern Hemisphere by SuperDARN. The time of year and time of day of the observations, and the IMF orientation and strength, are similar to the period presented by *Milan et al.* [2000]. However, the nature of the FTE signatures differs considerably, being confined to a convection throat not much more than 2 h of MLT in width. We investigate the possible reasons for the observed differences between the two intervals and discuss the ramifications for solar wind-magnetosphere coupling.

## 2. Observations

We compare the auroral signatures of FTEs from 26 August 1998, previously reported by *Milan et al.* [2000], with new observations from 31 August 2005.

The observations of *Milan et al.* [2000] are summarized in Figure 1. The solar wind and IMF conditions for the period, 09–13 UT 26 August 1998, are indicated in Figure 1k. There was a southward turning of the IMF shortly before 10 UT, after which, for the subsequent hour or so, IMF $B_Z \approx -10$ nT, $B_Y \approx 9$ nT, and the solar wind speed $V_X \approx 650$ km s$^{-1}$. Figure 1i shows the auroral evolution along the 16 MLT meridian as measured by the UVI camera on board Polar. Between 09 and 10 UT, when the IMF was directed northward, the auroral oval was static, located near a magnetic latitude of 76°. After the southward turning of the IMF the auroras progressed to lower latitudes as the polar cap expanded owing to dayside reconnection, the opening of magnetic flux, and the expansion of the polar cap, reaching 66° by 11 UT. From the observed rate of expansion of the polar cap during this period, *Milan et al.* [2000] estimated a reconnection rate close to 160 kV. The auroras were not seen between 11:00 and 11:30 UT due to the UVI camera being repointed. However, by 11:30 UT the auroras had retreated poleward again to a latitude of 73°, due to the onset of a substorm, the closure of magnetic flux, and thus the contraction of the polar cap. Although they are not very clear in this panel due to the relatively low cadence of images (≈3 min), a sequence of poleward moving auroral forms (PMAFs), which *Milan et al.* [2000] identified as the signatures of FTEs, was observed after 10 UT; the most clear examples are highlighted by arrows.

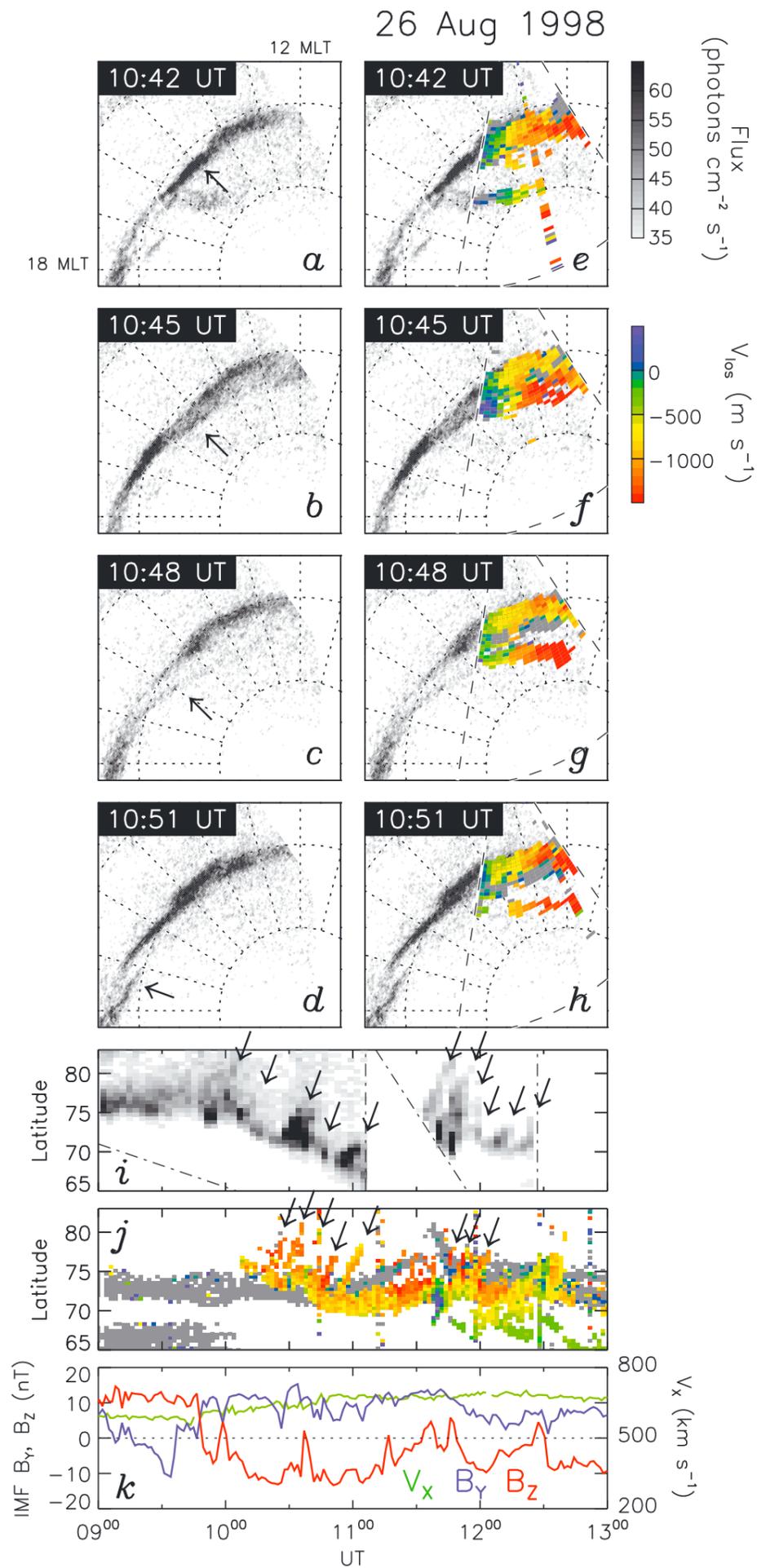

**Figure 1.** (a–d) A sequence of four auroral images taken by the UVI camera on board Polar on 26 August 1998, presented in a magnetic latitude and MLT frame, where concentric circles represent latitudes of 60, 70, and 80°. Arrows indicate the auroral feature that brightens and develops into a poleward moving auroral form (PMAF). (e–h) The same sequence of images with Hankasalmi SuperDARN radar observations of the line of sight motion of the ionosphere superimposed. (i) The auroral evolution along the 16 MLT meridian between 09 and 13 UT. The dot-dashed line indicates the edge of the field of view of the camera (the camera was repointed shortly after 11 UT). The most obvious poleward moving forms are highlighted by arrows. (j) Line of sight Doppler velocities along beam 7 (center of the field of view) of the Hankasalmi radar. Grey indicates ground scatter. Several poleward moving forms are highlighted by arrows. (k) IMF $B_Y$ and $B_Z$ and solar wind speed measured by Wind, propagated to the nose of the magnetosphere. After *Milan et al.* [2000].

Figure 1j shows line of sight velocity measurements along beam 7 of the Hankasalmi SuperDARN radar (pointing roughly along the 13 MLT meridian). Ionospheric backscatter was observed shortly after the southward turning of the IMF, at 10:10 UT. Ionospheric backscatter is seen where plasma irregularities are present in the F region ionosphere which meet the Bragg-scatter criterion with the wavelength of the radar signals (refer to *Milan et al.* [1997] for a detailed discussion of the conditions that lead to backscatter). The region of backscatter was closely associated with the location of the auroral oval, where irregularities are expected to be generated [*Milan et al.*, 1998]. This region moved equatorward until 11 UT, retreated poleward until 11:50 UT and then moved equatorward again, mirroring the behavior of the auroras seen by UVI (the latitudes differ because Figures 1i and 1j show observations from different MLT meridians). The red color coding of the backscatter indicates poleward line of sight flows into the polar cap with speeds near 1 km s⁻¹. A sequence of poleward moving backscatter features was seen, highlighted by arrows. These are the radar equivalent of the PMAFs seen by UVI [e.g., *Provan et al.*, 1998], sometimes known as poleward moving radar auroral forms (PMRAFs) [e.g., *Wild et al.*, 2001]. An exact one-to-one correspondence between PMAFs and PMRAFs is not clear due to the differing spatial and temporal resolutions of the measurements. This correspondence is investigated in Figures 1a–1h.

Figures 1a–1d show a sequence of four UVI auroral images in which one of the PMAFs is seen to develop. The auroral oval is seen to brighten in the 15 MLT sector, split, and an auroral feature (the PMAF) move into the polar cap. This motion corresponds to poleward convection flow associated with magnetopause reconnection. When observations from the Hankasalmi radar are superimposed (Figures 1e–1h), which covers the 12–14 MLT region, it is clear that the ionospheric feature extends across at least 7 h of MLT. As discussed by *Milan et al.* [2000], the pattern of line of sight velocities observed by the radar indicate poleward and dawnward flows (of up to 1.6 km s⁻¹), consistent with the expected sense of the magnetic tension force on newly opened magnetic field lines for $B_Y > 0$ nT [e.g., *Cowley et al.*, 1991; *Ruohoniemi and Greenwald*, 1996]. The flow and auroral dynamics associated with each FTE were interpreted by *Milan et al.* [2000] in terms of a reconnection X line that extended across a large fraction of the dayside magnetopause, but which was first active near noon, with activity then extending toward the flanks. Similar observations by *McWilliams et al.* [2001] supported this interpretation.

The new period of interest is 11–14 UT, 31 August 2005. The IMAGE FUV/WIC instrument observed 120–140 nm auroral emissions from the Southern Hemisphere, while SuperDARN radars were measuring the ionospheric convection flows in the Northern Hemisphere. Figure 2 shows observations from 12:38 UT, presented in a geomagnetic latitude and magnetic local time (MLT) coordinate system. Figure 2a shows the fields of view of the radars which observed ionospheric backscatter. To aid discussion below, two of the radars have particular beams highlighted in red, beams 6 and 12 of the Hankasalmi and Stokkseyri radars, as observations from these will be examined in detail below. These beams are indicated in Figures 2b and 2c also.

Figure 2b presents the Northern Hemisphere ionospheric flows deduced from the SuperDARN observations. Colored circles indicate where backscatter was observed by the radars, while the black contours are the electrostatic potential pattern—or, equivalently, ionospheric flow streamlines–inferred using the "map potential" technique described by *Ruohoniemi and Baker* [1998]. The colors represent the speed of the ionospheric flows, and the directions of the vectors indicate the flow direction. The electrostatic potential contours are separated by 6 kV, and the pattern as a whole had a cross-polar cap potential of 78 kV. As will be discussed below, the IMF had components $B_Y \approx 15$ nT and $B_Z \approx -8$ nT at this time. The convection pattern is consistent with expectations of a period of $B_Y > 0$ nT, southward IMF: a well-developed twin-cell pattern with dawnward directed flow in the dayside convection throat associated with magnetic tension forces on newly reconnected field lines and a well-defined convection reversal boundary in the dawn sector. The flows in the throat region approach 1.5 km s⁻¹ at this time.

Figure 2c shows the simultaneous Southern Hemisphere auroral observations from IMAGE FUV/WIC. Blue indicates dim or no emission and red bright emission (on an arbitrary scale). The observations have been mirrored about the noon-midnight meridian, so that dawn and dusk sector auroral emissions appear in the dusk and dawn sectors in Figure 2c, respectively. We do this because $B_Y$-associated asymmetries in the convection and auroral patterns are expected to be oppositely directed in the Northern and Southern hemispheres. By mirroring the auroral observations in this way, we can directly compare the observations from the Northern and Southern Hemispheres. Henceforth, unless otherwise stated, we will discuss the MLT location of auroral features as if they are mapped to the Northern Hemisphere in this way.

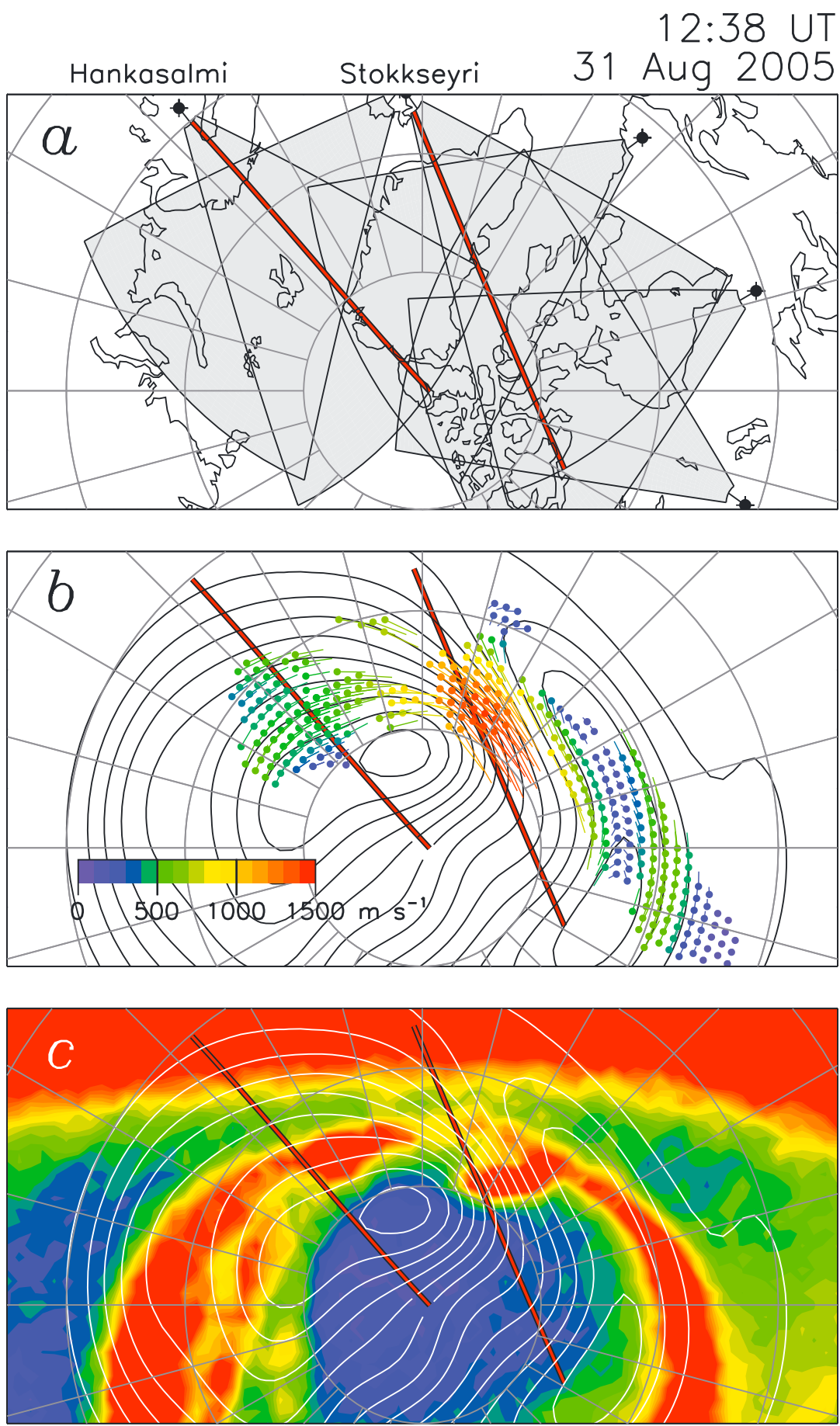

**Figure 2.** (a) The fields of view of SuperDARN radars at 12:38 UT on 31 August 2005, presented in a magnetic latitude and MLT frame. Beams 6 and 12 of the Hankasalmi and Stokkseyri radars are highlighted in red. (b) The SuperDARN convection measurements indicated by dots and tails, color coded by convection speed, where backscatter was observed, and the flow streamlines or electrostatic potential contours (separated by 6 kV). (c) The simultaneous auroral observations by IMAGE FUV/WIC on an arbitrary scale (red indicated bright features). Dayglow is apparent at the top of the panel.

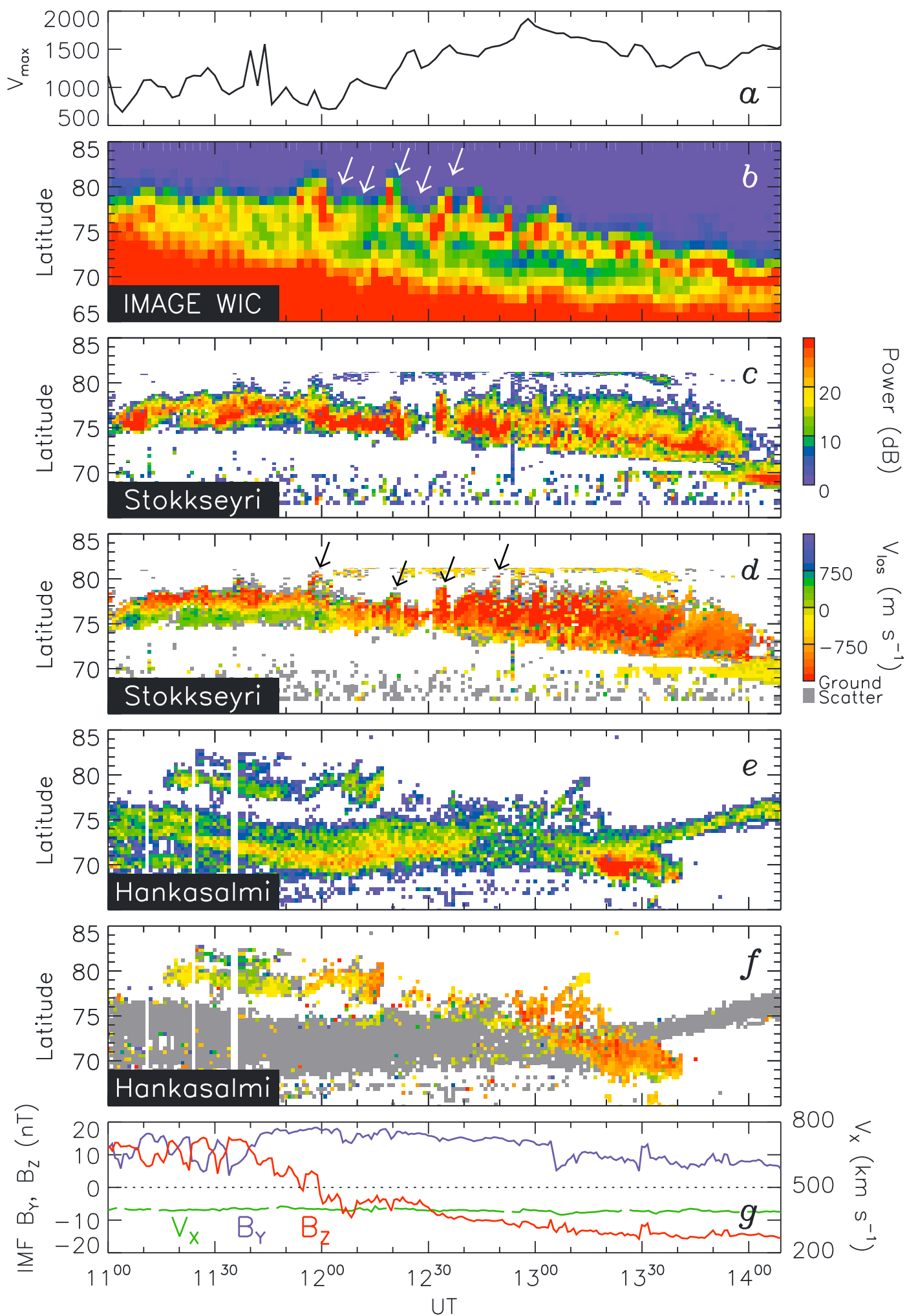

**Figure 3.** (a) The maximum ionospheric drift velocity within the dayside convection throat, 11–14 UT, 31 August 2005. (b) The auroral luminosity measured by IMAGE FUV/WIC along the position of beam 12 of the Stokkseyri radar (see Figure 2). Dayglow is apparent at the bottom of the panel. White arrows indicate the times of PMAFs identified in Figure 4. (c and d) The backscatter power and line of sight ionospheric drift velocity along beam 12 of the Stokkseyri radar. Black arrows indicate the times discussed in the text. (e and f) The backscatter power and line of sight ionospheric velocity along beam 6 of the Hankasalmi radar. (g) IMF $B_Y$ and $B_Z$ and solar wind speed, $V_X$ from the OMNI data set.

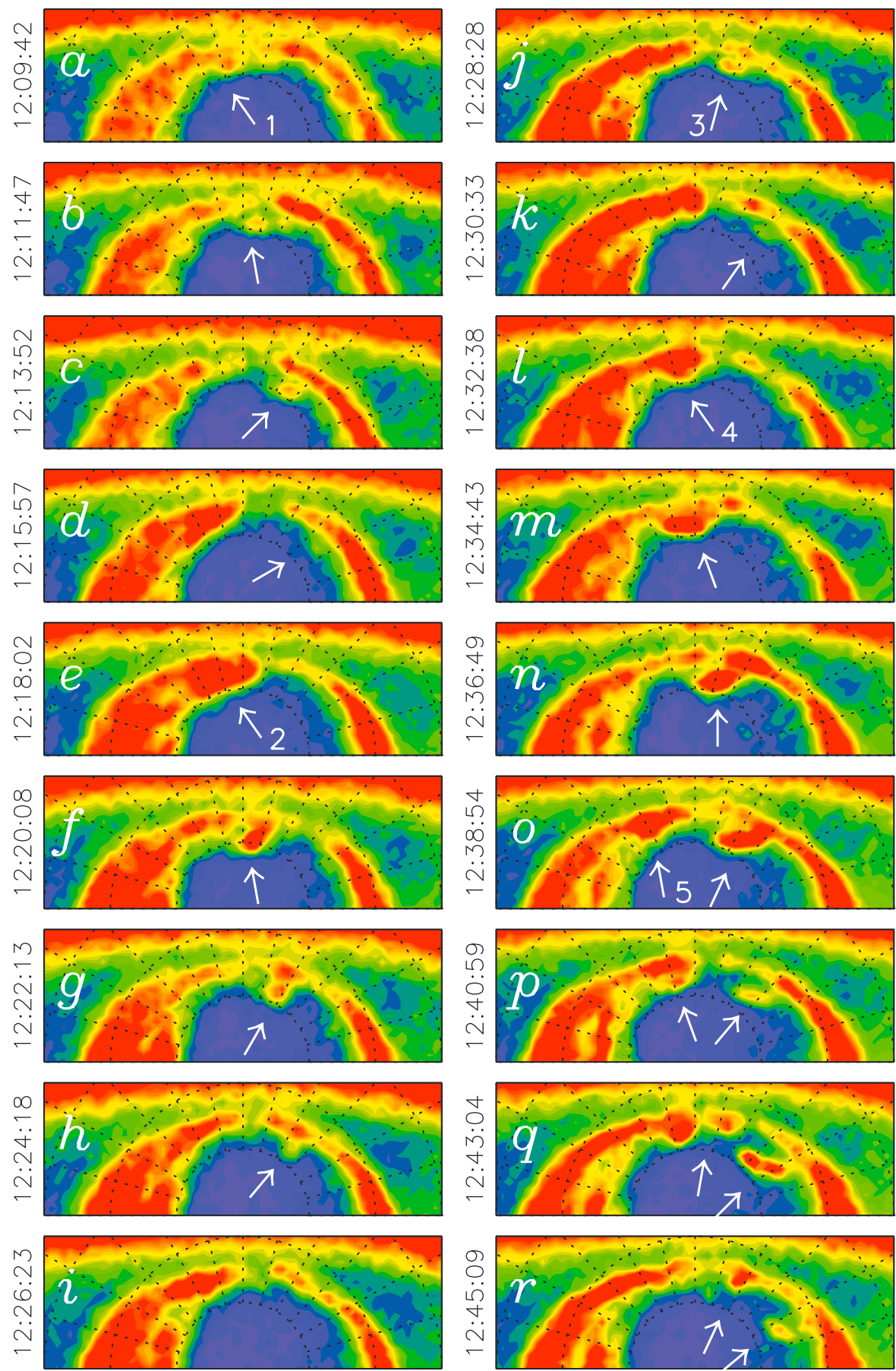

**Figure 4.** (a) A sequence of auroral images from IMAGE FUV/WIC between 12:09 and 12:45, 31 August 2005. The five features identified as PMAFs are indicated by arrows, numbered the first time a new feature appears.

A region of dayglow is visible at the top of Figure 2c. Also apparent is the auroral oval. The dawn sector oval is relatively narrow and is located near the equatorward side of the dawn sector convection reversal boundary. The dusk sector oval is thicker and more structured, associated with a transpolar arc that formed during the preceding period of northward IMF. A bright lozenge-shaped feature lies just poleward of the main oval, between 09 and 11 MLT. As will be discussed below, we interpret this as the auroral signature of a flux transfer event. Note that this feature sits within—and is of a similar width as—the ionospheric convection throat observed in the Northern Hemisphere.

Figure 3 presents the same data as a time series. Figure 3g shows the $B_Y$ and $B_Z$ components of the IMF and the solar wind speed $V_X$ (from the OMNI data set) [*King and Papitashvili*, 2005] during the period 11–14 UT, of which 12–13 UT is of most interest. Throughout the interval, $V_X \approx 380$ km s$^{-1}$. At 12 UT the IMF turns from northward to southward, varying from $-5$ nT to $-11$ nT over the next hour. IMF $B_Y \approx 17$ nT throughout this hourlong period.

Figure 3b shows the WIC auroral evolution along beam 12 of the Stokkseyri radar. At the bottom of the panel, mainly below 70° magnetic latitude, is a region of dayglow, which retreats to lower magnetic latitudes as the period progresses. The dayside oval is seen between 75° and 79° at the start of the interval. It remains near this latitude until about 12:15 UT, around the time of the southward turning of the IMF, and then it gradually progresses to a latitude of 70° by 14 UT. This movement to lower latitudes indicates that the polar cap expands due to reconnection occurring at the low-latitude dayside magnetopause associated with IMF $B_Z < 0$ nT. Between 12 and 13 UT there are brightenings of the dayside auroras that subsequently progress poleward, especially seen to start near 11:55, 12:15, 12:30, 12:40, 12:47, and 12:55 UT. We identify these as poleward moving auroral forms (PMAFs). Five arrows mark events which will be discussed below; some of these do not have a significant auroral signature in this figure.

Figures 3c and 3d show backscatter along beam 12 of the Stokkseyri radar. As in Figures 1i and 1j, there is a close correspondence between the locations of auroral features and radar backscatter. Figure 3c shows the power of the backscatter returns, whereas Figure 3d shows the line of sight motion of the ionosphere.

At the start of the interval, backscatter is observed between 75 and 79°, coincident with the auroral zone as seen in the WIC observations. Between 11 and 12 UT, the line of sight velocities show a flow shear (positive/negative velocities are toward/away from the radar), whose location is coincident with the expected location of the convection reversal boundary in the dawn convection cell. After 12 UT the velocities become negative, consistent with poleward flows into the polar cap in the dayside convection throat. After this time, the backscatter progresses equatorward, mirroring the expansion to lower latitudes of the auroral zone seen in the WIC observations.

Between 11:55 and 13:30 UT, the radar observes enhancements in backscatter power and enhancements in Doppler shift; some of which had a poleward progression along the radar beam (though the beam did not point directly meridionally), which we identify as PMRAFs. Some of these features appear to be associated with the PMAFs observed by WIC, especially at 11:55, 12:20, 12:32, and 12:50 UT (times indicated by arrows).

Figures 3e and 3f show backscatter power and Doppler shift from beam 6 of the Hankasalmi radar. The backscatter is scrappier in this case, but the equatorward progression of the backscatter is again apparent, and there is evidence for PMRAFs in the backscatter near 11:55, 12:15, and between 12:50 and 13:30 UT.

Finally, Figure 3a shows the convection speed within the convection throat, which varies between 1 and 1.5 km s$^{-1}$ during the main period of interest.

Figure 4 presents a sequence of WIC images between 12:09 and 12:45 UT, the period during which clear PMAFs were observed in Figure 3b. Each panel concentrates on the dayside auroras, with noon toward the top. Of interest to the present study is a sequence of five auroral features that appeared near 13 MLT (as viewed in the Northern Hemisphere) and progressed around the edge of the polar cap to earlier MLTs (in the Northern Hemisphere). Each feature is highlighted by an arrow; each arrow is numbered the first time the corresponding feature is clearly observed. The most conspicuous event, number 4, is that presented in Figure 2c. All features follow the same path in the ionosphere, closely aligned with the convection throat seen in Figure 2b.

## 3. Discussion

We have presented new observations of the Southern Hemisphere dayside auroral dynamics and the Northern Hemisphere ionospheric convection flows during a period of southward IMF, $B_Z \approx -7$ nT, around 12 UT on 31 August 2005, when significant low-latitude magnetopause reconnection is ongoing. The northern convection throat at this time is approximately 2 h of MLT in width, with a cross-throat voltage near 78 kV. In the Southern Hemisphere, a sequence of poleward moving auroral forms (PMAFs) is observed, which we interpret as the auroral signature of episodic magnetopause reconnection of flux transfer events (FTEs).

The IMF has a significant east-west component, $B_Y \approx 17$ nT, such that magnetic tension forces on newly opened field lines are expected to produce dawnward and duskward motions in the Northern and

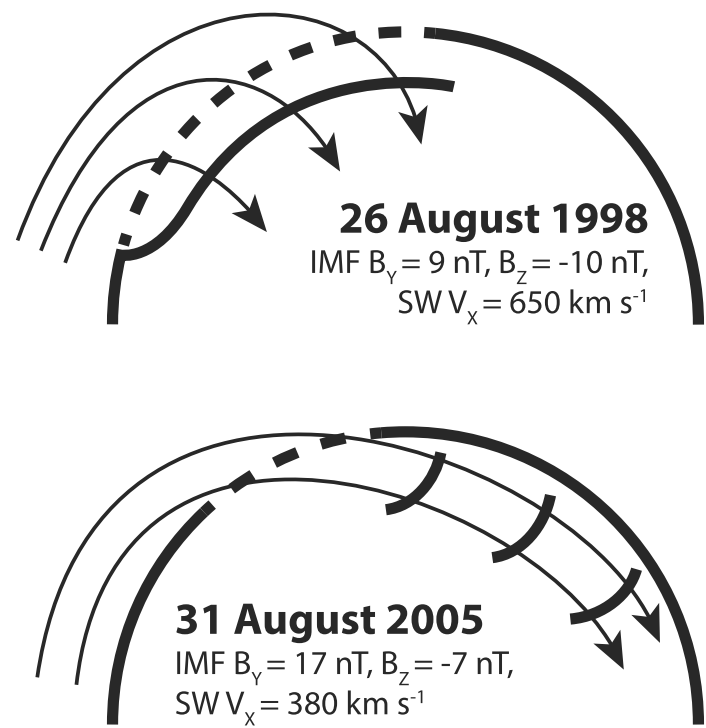

**Figure 5.** A schematic of the convection flow and auroral dynamics on 26 August 1998 and 31 August 2005. The semicircles represent the dayside polar cap boundary; the dashed portion of which is the ionospheric projection of the reconnection X line. The arrowed lines are convection streamlines. Features within the convection flow represent poleward moving auroral forms.

Southern Hemispheres, respectively. The Northern Hemisphere convection is consistent with this, as is the motion of the PMAFs in the Southern Hemisphere. Mirroring the southern auroral observations about the noon-midnight meridian places the path of the PMAFs within the northern convection throat, suggesting that the response to reconnection in the two hemispheres is highly symmetrical, though with the east-west sense reversed.

Observations around 10 UT on 26 August 1998 (almost exactly the same time of year) under similar IMF conditions, $B_Y \approx 9$ nT and $B_Z \approx -10$ nT, show transient reconnection signatures across 7 h of MLT and an estimated reconnection voltage up to 160 kV [Milan et al., 2000]. The main difference between these events was the solar wind speed, $V_X \approx 380$ km s$^{-1}$ on 31 August 2005, and $V_X \approx 650$ km s$^{-1}$ on 26 August 1998. The observations during the two intervals are summarized schematically in Figure 5. We note in passing that the observations of small-scale FTE features by Oksavik et al. [2005] were made during a period of 450 km s$^{-1}$ solar wind speed.

From observations of the rate of expansion of the polar cap during substorm growth phases, Milan et al. [2012] determined a formulation for the dayside reconnection rate in terms of upstream solar wind conditions of

$$\Phi_D = \Lambda V_X^{4/3} \sqrt{B_Y^2 + B_Z^2} \sin^{9/2} \tfrac{1}{2} |\theta|, \tag{1}$$

where $\Lambda = 3.3 \times 10^5$ m$^{2/3}$ s$^{1/3}$, $\theta$ is the IMF clock angle, $V_X$ is expressed in m s$^{-1}$, and $B_Y$ and $B_Z$ in T. This equation yields average values of $\Phi_D \approx 75$ kV and 165 kV for the 31 August 2005 and 26 August 1998 cases, respectively. These predictions are close to the observed cross-throat voltage and deduced reconnection rate for these two intervals.

Milan et al. [2000] interpreted the observations from 26 August 1998 as indicating that in each reconnection event (FTE) the X line developed first near noon and then extended antisunward across the magnetopause, with a total width of 7 h of MLT. It was estimated that the instantaneous length of the reconnection X line was on average 4 h of MLT during this period. In the 31 August 2005 case, we do not see a similar temporal development, but a convection throat with a fixed width of 2 h of MLT. In both cases, the ionospheric convection speed within the convection throat was similar, close to 1.5 km s$^{-1}$.

It appears, then, that the greater reconnection rate occurring on 26 August 1998, mainly associated with a faster solar wind, is accommodated by having a longer reconnection X line on the magnetopause, rather than a higher rate of reconnection in the same limited MLT extent. We conclude that the length of the magnetopause reconnection X line is controlled by the speed of the solar wind. The dynamics of the X line, propagating or static, also appears to be dependent on solar wind speed or the overall reconnection rate.

Although the quasiperiodic nature of FTEs is well known, there is no satisfactory mechanism that has been proposed to explain it. Observations at the magnetopause suggest that interevent periods have a broad but skewed distribution with lower and upper deciles of 1.5 and 18.5 min, a mode of 3 min, and a mean of 8 min [*Lockwood and Wild*, 1993]. In the case of *Milan et al.* [2000], five FTEs were reported over an interval of 50 min, an interevent duration close to 10 min. In the present example, Figure 4 shows five FTEs over an interval of 35 min, an interevent duration near 7 min. It is unclear what determines this repetition rate, but in these cases it appears that a higher overall reconnection rate is achieved by larger, longer duration events rather than more events of the same duration or magnetic flux content.

## 4. Conclusions

Observations of the auroral and convection signatures of flux transfer events at two similar dates and times and under similar IMF strengths and orientations, but differing solar wind speed, show significant differences. Higher (lower) solar wind speeds result in a higher integrated reconnection rate, larger (smaller) flux content FTEs, a wider (narrower) extent of reconnection on the magnetopause and ionospheric convection throat, a slower (faster) event repetition rate, and a propagating (static) X line.

These differences may play a role in controlling the magnetopause reconnection rate for different solar wind conditions. Investigating the cause of these differences is essential for understanding the solar wind driving of the magnetosphere. Clearly, a systematic study of the extent and variability of ionospheric features associated with magnetopause reconnection under a variety of solar wind conditions is warranted.

**Acknowledgments**
S.E.M. and J.A.C. were supported by the Science and Technology Facilities Council (STFC), UK, grant ST/K001000/1. M.T.W. was supported by a STFC studentship. This study was also supported by the Research Council of Norway under contract 223252/F50. The OMNI data were obtained from the GSFC/SPDF OMNIWeb interface at http://omniweb.gsfc.nasa.gov. The SuperDARN data are available from the Virginia Tech SuperDARN data portal at http://vt.superdarn.org. Polar UVI data were downloaded from the NASA/MSFC Space Plasma Physics data portal at http://spacephysics.msfc.nasa.gov/ projects/uvi/data_archives.shtml. IMAGE FUV/WIC data were accessed from the University of California at Berkeley data center at http://sprg.ssl.berkeley.edu/image/.